\documentclass[a4paper]{spie}  %>>> use this instead for A4 paper
\usepackage{amsmath,amsfonts,amssymb}
\usepackage{graphicx}
\usepackage[colorlinks=true, allcolors=blue]{hyperref}
\usepackage{enumitem}
\setlist{nosep} % or \setlist{noitemsep} to leave space around whole list

\title{The MICADO first light imager for the ELT:\\Sparse Aperture Masks, design and simulations}

\author[a]{E. Huby}
\author[a]{P. Baudoz}
\author[a]{S. Lacour}
\author[a]{M. Le Teuff}
\author[a]{Y. Cl\'enet}
\author[b]{R. Davies}
\affil[a]{LESIA, Observatoire de Paris, Universit\'e PSL, CNRS, Sorbonne Universit\'e, Universit\'e Paris Cit\'e, 5 place Jules Janssen, 92195 Meudon, France}
\affil[b]{Max-Planck-Institut für extraterrestrische Physik, Garching, Germany}

\authorinfo{Send correspondence to E. H.: elsa.huby@obspm.fr}

\begin{document} 
\maketitle

\begin{abstract}
MICADO, the European Extremely Large Telescope first light imager will feature a dedicated high contrast imaging mode specifically designed for observing and characterizing exoplanets and circumstellar disks. Two sparse aperture masks (SAM) will be included, consisting of opaque masks with a set of holes arranged in a non-redundant configuration. Pupil masking transforms a monolithic telescope into an interferometer, with the aim of recovering spatial information down to the diffraction limit of the telescope and below, even in the presence of residual aberrations, such as turbulent AO residuals, and non common path aberrations. On the ELT, SAM will enable the detection of features down to 3.3\,mas in the J band, 12\,mas in the K band. Two designs have been chosen and will be reviewed, with complementarity in terms of sensitivity and spatial frequency coverage for image reconstruction. In this contribution, the technical choices will be detailed, such as the hole diameter and arrangement, given the technical constraints such as spectral filter bandwidths, detector dimensions, sampling, read-out-noise and frame rate. We will also report on simulations performed to assess the expected capabilities of this mode, with application examples of close companion detection and contrast curves.
\end{abstract}

\keywords{ELT/MICADO, Sparse Aperture Masking, Non redundant imaging, high angular resolution}

\section{INTRODUCTION}
\label{sec:intro} 

 The MICADO instrument\cite{Davies2016} will offer a high contrast mode, including the following observing modes:
\begin{itemize}
    \item classical Lyot coronagraphs, consisting of the combination of an occulting mask in the focal plane and a Lyot stop in the pupil plane\cite{Perrot2018}. There will be three available occulting masks with different radii. The mask prototypes are currently being tested and characterized\cite{Baudoz2024}.
    \item Vector Apodized Phase Plate (VAPP), a phase mask located in a pupil plane, which reshapes the incoming beam diffracted in the subsequent focal plane to carve a dark region around the central star, where the starlight is rejected elsewhere in the field of view\cite{Codona2006,Snik2012,Doelman2021}.
    \item Sparse Aperture Masks (SAM): there are two available masks with non-redundant configurations of 9 and 18 holes, whose design will be described in the following sections. 
\end{itemize}

Sparse aperture masking has proven to be powerful in restoring performance at the diffraction limit of telescopes, and even below, from the ground\cite{Haniff1987, Tuthill2000, Lacour2011} and from space onboard the JWST\cite{Sallum2024}. In particular, measuring the closure phase\cite{Jennison1958}, consisting of combining the interferometric phase of three baselines forming a closed triangle, provides an observable that is robust to differential piston terms that may remain between the sub-apertures. As such, sparse aperture masking is a technique that can be used even under turbulent conditions, if the hole diameters are designed to match the typical width of a coherence patch of the turbulence, i.e. the Fried parameter. This feature makes it a powerful technique to reach very high angular resolution.

Using an aperture mask downstream of an AO system does not cancel these advantages, but rather enhances its sensitivity. Indeed, the AO system provides a correction of the wavefront, reducing the residual piston terms between the sub-apertures, thus allowing for longer exposure times, while closure phases handle the residual piston terms. In addition, high order aberrations are also corrected across the single sub-apertures, resulting in increased coherent patches and allowing for larger diameter holes.

In this paper, the method for defining the chosen non-redundant configurations is described in Section\,\ref{sec:config}. The study carried out on the impact of the holes' diameter on throughput and closure phase error is detailed in Section\,\ref{sec:opti}. Finally, the results of a simulated detection of a faint close-in companion with the selected 9-hole mask are reported in Section\,\ref{sec:simu}.

\section{Non redundant configurations: 9-hole and 18-hole masks}
\label{sec:config}

\begin{figure}
    \centering
    \includegraphics[width=.6\linewidth]{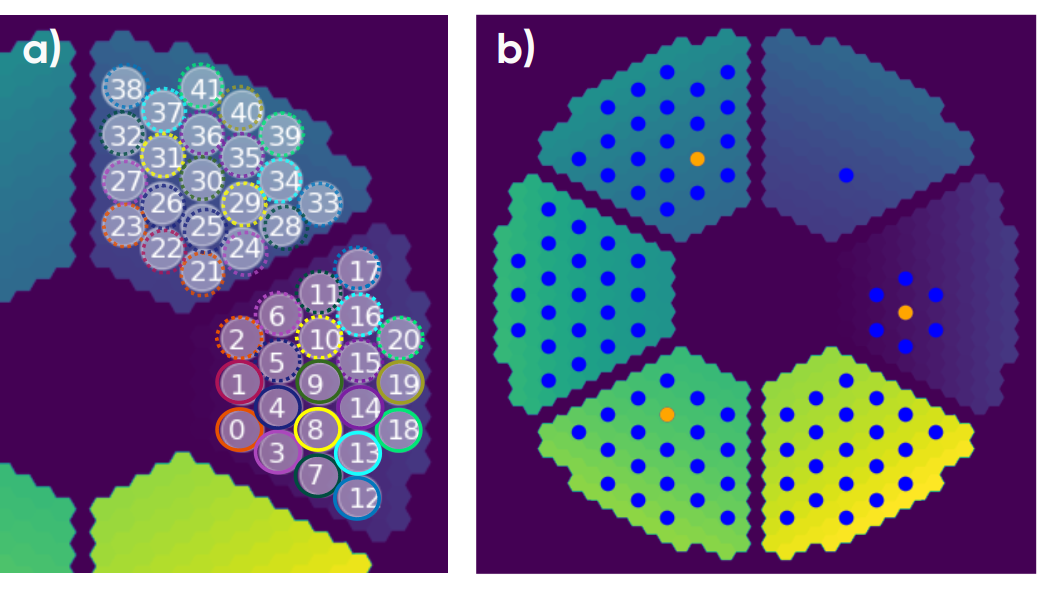}
    \caption{Search grid for placing the holes. a) The 21 possible locations on a single fragment of the ELT pupil, leaving a space of 2 segments in between two locations. b) }
    \label{fig:grid}
\end{figure}

\paragraph{Hole search grid.} Initially, holes of the same diameter as the ELT segments were considered\cite{Lacour2014}. This led to holes of 1.17\,m in diameter, which produced a PSF that spread over too many pixels compared to the pixel number allowed in the fast read mode of MICADO's detector. Indeed a 10\,Hz reading rate of the detector would only be possible by reducing the number of lines to be read to 300 (keeping 4096 pixels in the other direction). The pixel scale options will be 1.5 and 4\,mas per pixel, which limits the SAM field of view to 450\,mas or to 1200\,mas with the coarser sampling. For instance, at 1.635\,$\mu m$, the central lobe of a single-hole PSF would spread over 703\,mas. This not only exceeds the allowed field of view, but also leads to a significant contribution of the detector read-out-noise due to the large number of pixels used for the interferogram acquisition. It was thus decided to increase the hole size. The grid for the hole locations was thus limited to 1 out of 2 ELT-segments, as illustrated in Fig.\,\ref{fig:grid}-a).

\paragraph{Non-redundant configurations.} The search for non-redundant configurations was performed using a hexagonal grid to place the holes. A symmetry of 120\,degrees was imposed for the design, to limit the number of possible non-redundant configurations, while leading to homogeneous spatial frequency coverage. All configurations were searched by placing one hole at a time, and flagging all the other locations to be excluded to avoid redundancy, as illustrated in Fig.\,\ref{fig:grid}-b). Additionally, the shortest baselines were not allowed, as they are usually poorly constrained. In order to list all possible configurations, this iterative search was performed starting with each of the 12 locations highlighted in Figure\,\ref{fig:grid}-a). Starting with any other of the 29 possible locations would result in configurations already listed, due to the symmetry. Using this search algorithm, configurations with 9, 12, 15 and 18 holes were found.

\begin{figure}
    \centering
    \includegraphics[width=.9\linewidth]{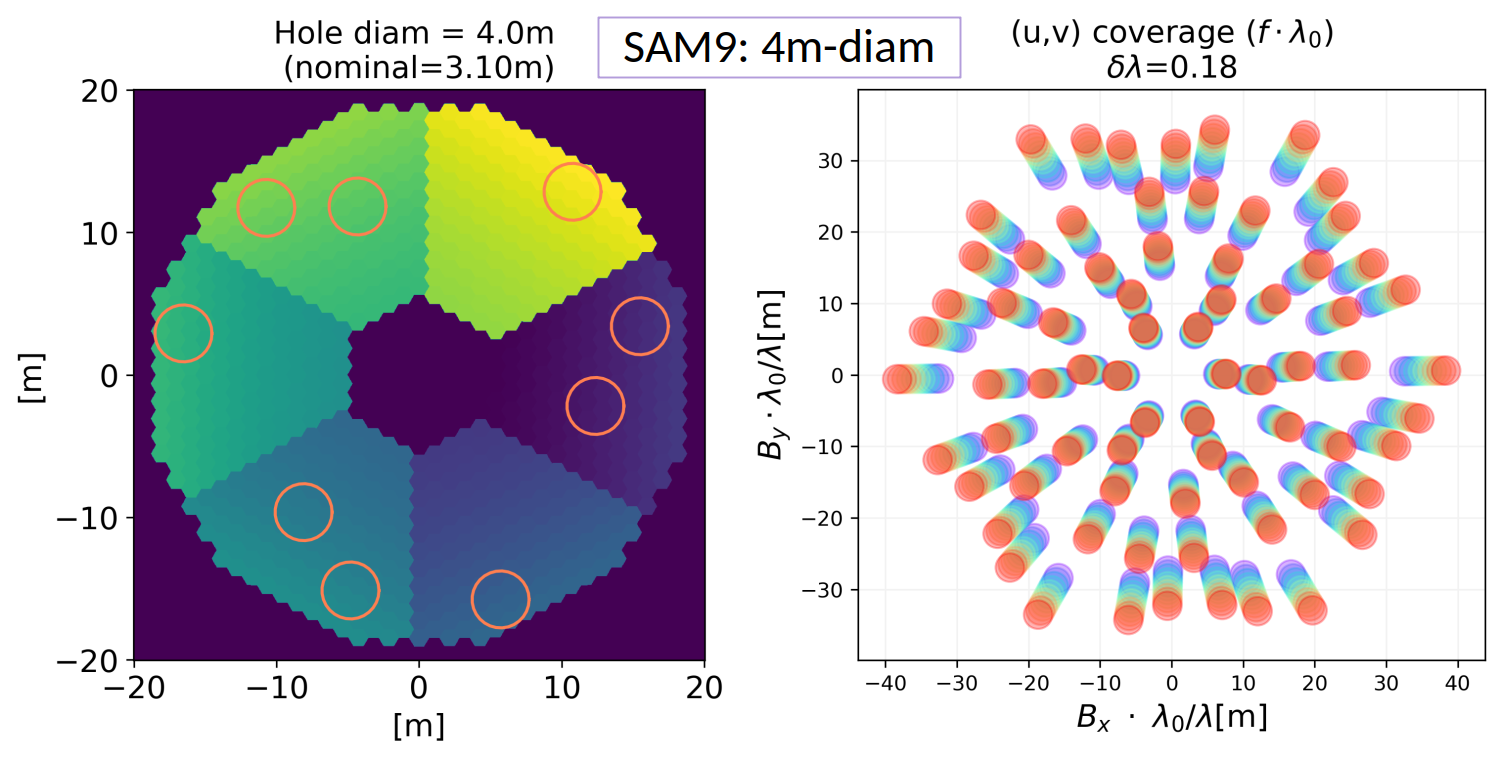}
    \caption{Selected 9-hole non-redundant configuration, with the corresponding (u,v) plane coverage corresponding to a spectral bandwidth of 18\%: blue and red spots represent the shortest and longest wavelengths, respectively. While the hole diameter avoiding any overlap of the frequency peaks corresponds to 3.10\,m, the configuration is here shown with 4m-diameter holes.}
    \label{fig:9hole_mask}
\end{figure}

\begin{figure}
    \centering
    \includegraphics[width=.9\linewidth]{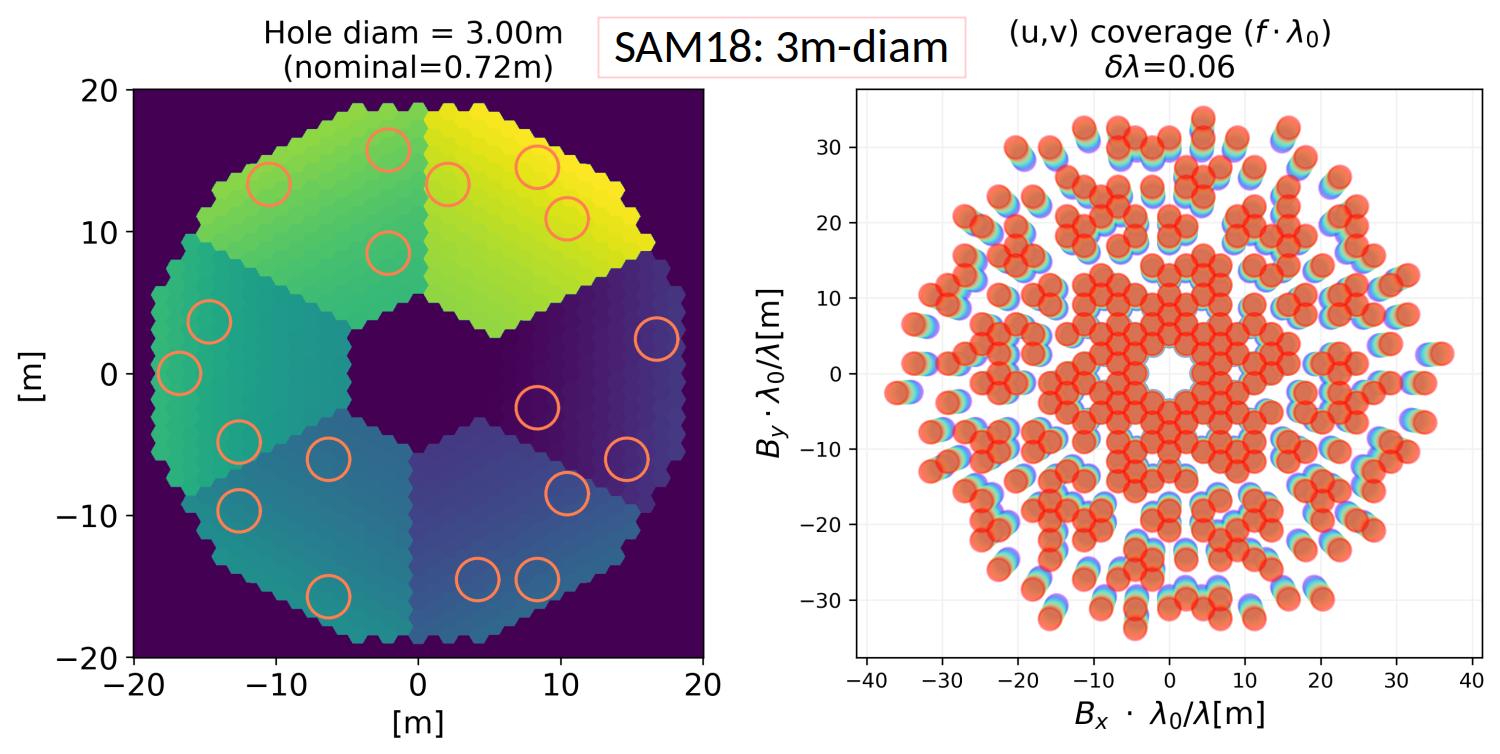}
    \caption{Same as Figure\,\ref{fig:9hole_mask} for the selected 18-hole non-redundant configuration, and a spectral bandwidth of 6\% for the (u,v) plane coverage. While the hole diameter avoiding any overlap of the frequency peaks corresponds to 0.72\,m, the configuration is here shown with 3m-diameter holes.}
    \label{fig:18hole_mask}
\end{figure}

\paragraph{Two configurations were chosen:}
\begin{itemize}
    \item a 9-hole mask, for high accuracy phase measurement, envisioned for searching companions at high contrast.
    \item a 18-hole mask, for image reconstruction, envisioned for imaging disks.
\end{itemize}

For the selected configurations, the hole locations were chosen among all the found configurations by considering the spacing between the spatial frequency peaks in the spectral domain. This spacing must be maximized to avoid the overlap of the different spatial frequency components. Each peak is located at coordinates corresponding to the spatial frequency $\boldsymbol{B} / \lambda$ and extends over an area of diameter $2 D_{\rm{hole}} / \lambda$ (the Fourier transform of the PSF of a single hole of diameter $D_{\rm{hole}}$). For every configuration, the minimal distance between frequency components is computed, taking into account a spectral bandwidth of 6\% or 18\%, corresponding to the typical width of the set of J, H and K filters that will be used in this observing mode. Following this criterion, the maximum diameter was found to be 3.10\,m for the 9-hole configuration and 0.72\,m for the 18-hole configuration. Fig.\,\ref{fig:9hole_mask} and \ref{fig:18hole_mask} show the chosen non redundant configurations for 9 and 18 holes, with 4m- and 3m-diameter holes, respectively. The following study aims at determining the impact of an increase in the hole diameter with respect to these limits.

\section{Hole diameter optimization}
\label{sec:opti}

Determining the hole diameter requires to take several parameters into account. Arguments towards large holes are the following:
\begin{itemize}
    \item an increased throughput ;
    \item a limited area for the interferogram, to fit inside the allowed field of view and limit the impact of the detector read-out-noise.
\end{itemize}
On the other hand, smaller holes are better suited for:
\begin{itemize}
    \item avoid spectral band pass smearing (overlap of the frequency peaks) ;
    \item robustness against the impact of aberrations across the sub-apertures.
\end{itemize}
The band pass smearing appears to be less critical than expected, as long as the spectral width is included in the model used to fit the fringes. In the simulated data, a basic flat spectrum was considered, obtained as the sum of 11 or 21 monochromatic components across the bandwidth for the narrow and broadband filters, respectively. However, difficulties might arise as soon as the unknown stellar spectrum differs from the flat spectrum used in the model. Further simulations are required to investigate about this impact.

The main trade-off that we considered was thus to increase the hole diameter as much as possible to maximize the global throughput, while maintaining a reasonable impact of the residual turbulence.

\subsection{Throughput}

\begin{figure}
    \centering
    \includegraphics[width=.9\linewidth]{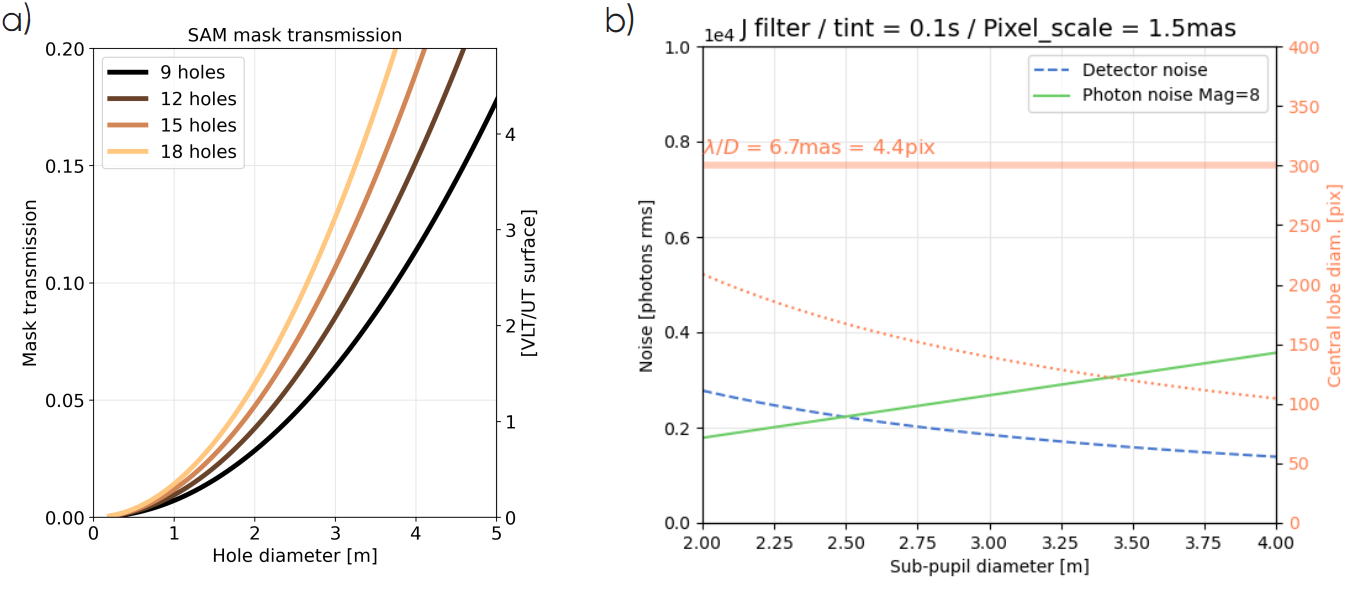}
    \caption{a) Global transmission of a mask with 9, 12, 15 or 18 holes, as a function of the hole diameter. The transmission is expressed as a pourcentage of the total ELT pupil area (left axis) and in unit of the surface of a single VLT/UT (right axis). b) Comparison of the noise contribution due to photon noise for an 8-magnitude target in broadband Ks-filter with a 9-hole mask and the detector noise, as a function of the hole diameter. The light orange curves refer to the right axis, representing the width of the central lobe of the PSF of a single hole. The horizontal line highlights the limit of the field of view at 300 pixels.}
    \label{fig:throughput}
\end{figure}

By definition, the throughput increases as the square of the hole diameter. The transmission of a sparse aperture mask compared to the ELT total collecting surface is shown in Fig.\,\ref{fig:throughput}-a). A 9-hole mask with 4m-diameter holes, and an 18-hole mask with 3m-diameter holes lead to a transmission of about 12\% in both cases. Given the tremendous diameter of the ELT of 39m, a transmission of 12\% is actually equivalent to the collecting surface of 3 Unit Telescopes of the VLT. In other terms, the MICADO sparse aperture masks would enable an angular resolution better than 10\,mas with a sensitivity equivalent to the collecting surface of 3 VLT/UT.

\subsection{Closure phase error}

\begin{figure}
    \centering
    \includegraphics[width=\linewidth]{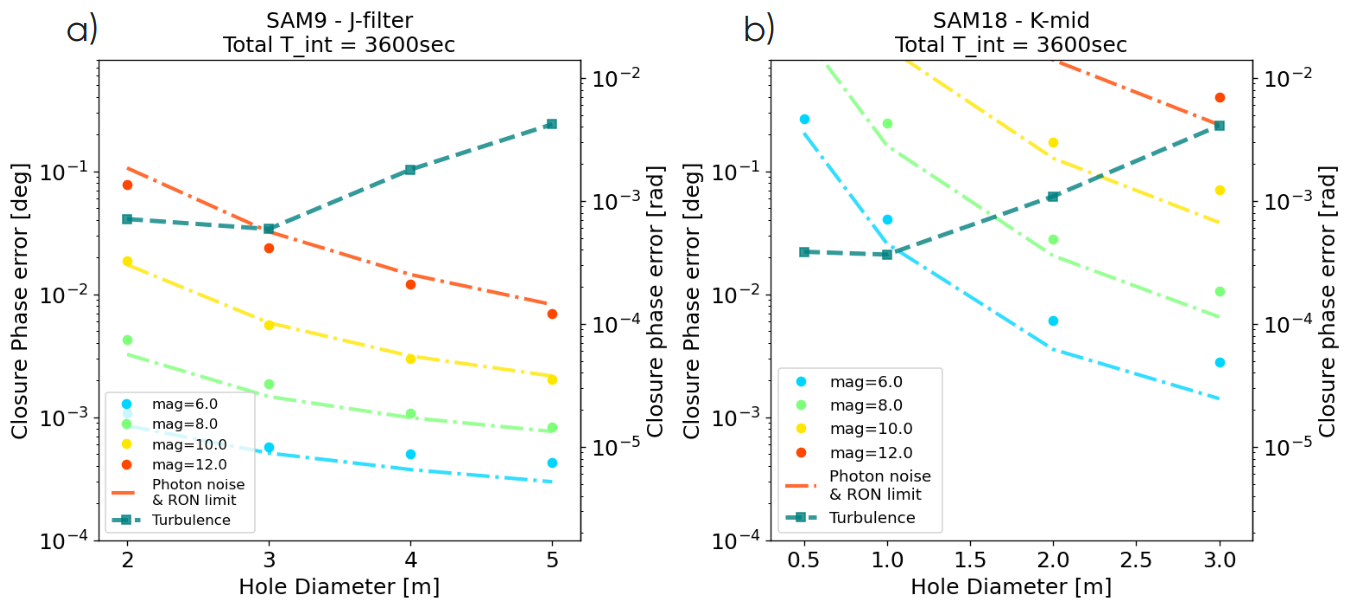}
    \caption{Closure phase error as a function of the hole diameter, obtained from simulated images with a) SAM9 in broadband J filter and b) SAM18 in K-mid filter, including read-out noise and photon noise for different magnitudes (color circles), compared with the theoretical limit (dot-dash line). The closure phase error obtained from interferograms only affected by residual turbulence (no photon noise) is also plotted.}
    \label{fig:cp_error}
\end{figure}

The standard data reduction process is based on the analysis of the closure phase\cite{Jennison1958, Haniff1987}, signals extracted from the interferogram, a quantity that is intrinsically calibrated against differential piston terms between remaining between the sub-apertures. The closure phase error is a proxy of the performance of this observing mode, as it is directly linked to the contrast performance when searching a faint companion\cite{Lacour2011}.

We have simulated cubes of images and reduced them using the standard P2VM method (Pixel to Visibility Matrix)\cite{Tatulli2007} to fit the fringes and estimate the complex coherent fluxes. Closure phases are estimated as the phase of the bispectrum. With 9 sub-apertures, 84 closure phases can be computed, while 816 closure phases are computed for an 18-hole mask. The final 84 or 816 closure phases are computed as the temporal averages over the data cuves. Finally, the error on the closure phase is estimated as the standard deviation over the 84 or 816 closure phases estimates.

First, we studied the impact of read-out noise and photon noise. In absence of any aberrations, we made the assumption that images can be temporally binned to reduce the total number of frames to be simulated. While we assume a typical integration time of 120\,ms, 9-sec exposure time images were simulated, with a read-out noise equivalent to a level of 15\,e- per 120\,ms-frames (that is 129.9\,e-). Cubes of 400 frames were simulated, representing 1\,hour of data. The error estimated by processing these simulated cubes are shown in Fig.\ref{fig:cp_error} in two cases: the 9-hole mask in broadband J filter and the 18-hole mask in K-mid filter. The theoretical limit as derived by Ireland\,2013\cite{Ireland2013} is also plotted and is consistent with the simulation results.

Independently, simulations were also run with turbulent AO residuals, computed with COMPASS\cite{Vidal2022}. There are 1300 phase screens sampled at 1\,Hz, correponding to median seeing conditions (0.70\,arcsec). They thus represent a sequence of 22\,min, which is shorter than the cubes simulated to estimate the photon noise impact. The results might thus be slightly pessimistic, but it appeared that the averaging process of the residual aberrations was not drastically changing the closure phase error when averaging over 600 or 1300 frames. The closure phase error induced by the AO residuals are also plotted in Fig.\,\ref{fig:cp_error} to directly compare its contribution to the photon noise.

The results shown in Fig\,\ref{fig:cp_error} show that for the 9-hole mask in broadband J filter, closure phase measurements with 4m-diameter holes would be limited by the AO residual aberrations, to about $2 \cdot 10^{-3}$\,rad, which is the current limit achieved by sparse aperture masking from the ground\cite{Stolker2024}. On the other hand, 3m-diameter holes for the 18-hole mask used with the K-mid narrow band filter would also be limited by the residual aberrations, at about $2 \cdot 10^{-3}$\,rad. This slight degradation in closure phase precision would be balanced by their large number, making this mask well suited for image reconstruction.

\section{Simulations: companion detection}
\label{sec:simu}

\begin{figure}
    \centering
    \includegraphics[width=\linewidth]{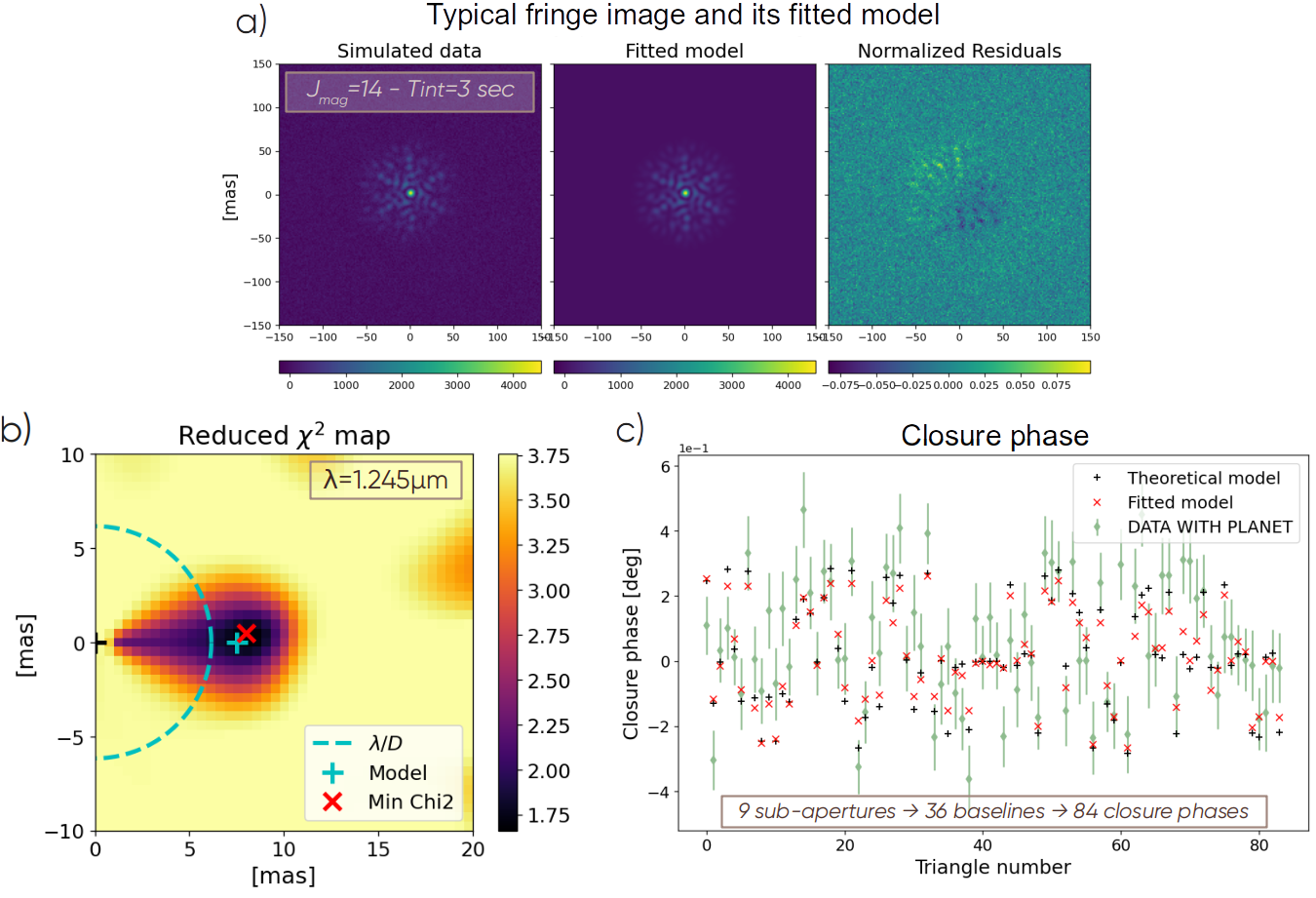}
    \caption{Simulation results of the detection of a companion around a $J_{\rm{mag}}$=14 star (broad band J filter), at a separation of 7.5\,mas or 1.1\,$\lambda/D$, and a flux ratio of $2 \cdot 10^{-3}$. a) One typical simulated frame (equivalent to 3\,sec integration time in terms of photon noise), the fitted model and the residuals. b) Reduced $\chi^2$ map (minimized along the flux ratio parameter axis) corresponding to the closure phase fitting. c) Estimated closure phase and best fit model from the $\chi^2$ minimization. Also plotted are the closure phases expected from the perfect model.}
    \label{fig:comp_det}
\end{figure}

The detection of a faint companion has been simulated to confirm the capabilities of this mode. An example of simulations is shown in Figure\,\ref{fig:comp_det}. A set of 1300 images were simulated, with turbulent AO residuals from the COMPASS simulation (equivalent to 20\,min in terms of diversity, see previous section), with the read-out noise and photon noise equivalent to about 1\,hour of observations (3\,sec integration time with 75\,e- rms of read-out-noise) for a target of magnitude 14 in broadband J. However, the sensitivity limit imposed by the exposure time that will be typically limited to 100\,ms has not been completely characterized yet. 

No other aberrations than the turbulent residual phase after AO correction has been included. It is indeed assumed that the impact of static non common path aberrations is calibrated out by the observation of a reference star. In the simulations, the tip-tilt component was removed from the AO phase screens, assuming that images are recentered during post-processing, prior to fringe fitting.

Closure phases are estimated from the phases obtained from fringe fitting of every single frame, thanks to the P2VM method\cite{Tatulli2007}. A binary system model is then fitted to the closure phases following a grid search to minimize the $\chi^2$ parameter. The parameters to fit are: (x,y), the coordinates of the separation vector of the system, and the flux ratio between the components. 

In the simulation example shown in Fig.\,\ref{fig:comp_det}, a companion that is $2 \cdot 10^{-3}$ fainter than the central star, and located at a separation 7.5\,mas, or 1.1\,$\lambda/D$, is recovered in the closure phase signal. The reduced $\chi^2$ map, obtained by minimizing the $\chi^2$ cube along the flux ratio parameter, is shown in Fig.\,\ref{fig:comp_det}-b. The 84 estimated closure phases are shown in Fig.\,\ref{fig:comp_det}-c, exhibiting a mean error bar of 0.1deg, along with the best fit model. 

\section{Conclusion}

The design of two sparse aperture masks is currently being finalized for future observations with the MICADO instrument. Non redundant imaging has proven to be powerful to detect features down to the diffraction limit of the telescope and below, even under residual turbulent phase perturbations. SAM observations are thus foreseen as a promising observing mode to achieve the ultimate resolution limit of the ELT, especially at the beginning of MICADO operations. 

Two masks will be offered: a 9-hole mask tailored for faint companion detection, providing accurate phase measurements, and an 18-hole mask, offering broader frequency coverage, suitable for image reconstruction of close-in disks, among other applications. A study of the optimal hole diameter has been conducted, showing that a trade-of has to be made between throughput and closure phase accuracy affected by residual turbulence perturbations. Configurations with larger hole diameters would provide SAM masks with a global transmission of approximately 12\% of the ELT surface area, equivalent to three times the collecting surface of a single 8m-telescope of the VLT observatory.

Basic companion detection simulations have been conducted for both masks. Further aspects will be explored in ongoing analysis, including defining sensitivity limits, simulating different spectral shape, assessing the effect of turbulent variations during single exposures, and evaluating the impact of non perfect closure phase calibration. Additionally, the capabilities in wavefront sensing, in particular for detecting petal modes\cite{Martinache2022, Deo2024}, will be investigated within the context of MICADO SAM mode.

\acknowledgments  
 
The MICADO project has benefited from the support of 1) the French Programme d’Investissement d’Avenir through the project F-CELT ANR-21-ESRE-0008, 2) the CNRS 80 PRIME program, 3) the CNRS INSU IR budget, 4) the Action Spécifique Haute Résolution Angulaire (ASHRA) of CNRS/INSU co-funded by CNES, 5) the Observatoire de Paris and 6) the Région Ile-de-France (DIM ACAV/ACAV+ and ORIGINES).

% References
\bibliography{spie_proc_sam} % bibliography data in report.bib
\bibliographystyle{spiebib} % makes bibtex use spiebib.bst

\end{document}